\documentclass[11pt,a4paper]{article}

\usepackage{jheppub}
\usepackage{subfig}
\usepackage{bbm}
\usepackage[T1]{fontenc}
\newcommand{\MSb}{\overline{\textrm{MS}}}
\usepackage{multirow} 

\title{
{\footnotesize\vspace*{-3.5cm}\hspace*{13cm}{\textnormal{DESY 13-038}}\vspace*{-0.4cm}}
{\footnotesize\hspace*{13cm}\textnormal{HU-EP-13/08}\vspace*{-0.4cm}}
{\footnotesize\hspace*{12.7cm}\textnormal{SFB-CPP-13-17}\vspace*{0.5cm}}
Chiral condensate from the twisted mass Dirac
operator spectrum}

\author[a,b]{Krzysztof Cichy,}
\author[a,c]{Elena Garcia-Ramos,}
\author[a,d]{Karl Jansen}

\affiliation[a]{NIC, DESY, Platanenallee 6, 15738 Zeuthen, Germany}
\affiliation[b]{Adam Mickiewicz University, Faculty of Physics,
Umultowska 85, 61-614 Poznan, Poland}
\affiliation[c]{Humboldt Universit\"at zu Berlin, Newtonstr. 15, 12489
  Berlin, Germany}
\affiliation[d]{Department of Physics, University of Cyprus, P.O. Box 20537, 1678 Nicosia,
Cyprus}

\emailAdd{krzysztof.cichy@desy.de}
\emailAdd{elena.garcia.ramos@desy.de}
\emailAdd{karl.jansen@desy.de}

\abstract{
We present the results of our computation of the dimensionless chiral condensate
$r_0\Sigma^{1/3}$ with $N_f=2$ and $N_f=2+1+1$ flavours of
maximally twisted mass fermions. The condensate is determined from the Dirac operator
spectrum, applying the spectral projector method proposed by Giusti and L\"uscher.
We use 3 lattice spacings and several quark masses at each lattice spacing to perform the
chiral and continuum extrapolations. We study the effect of the dynamical strange and charm quarks by
comparing our results for $N_f=2$ and $N_f=2+1+1$ dynamical flavours.
\begin{center}
\vspace*{1cm}
\includegraphics
[width=0.2\textwidth,angle=0]
{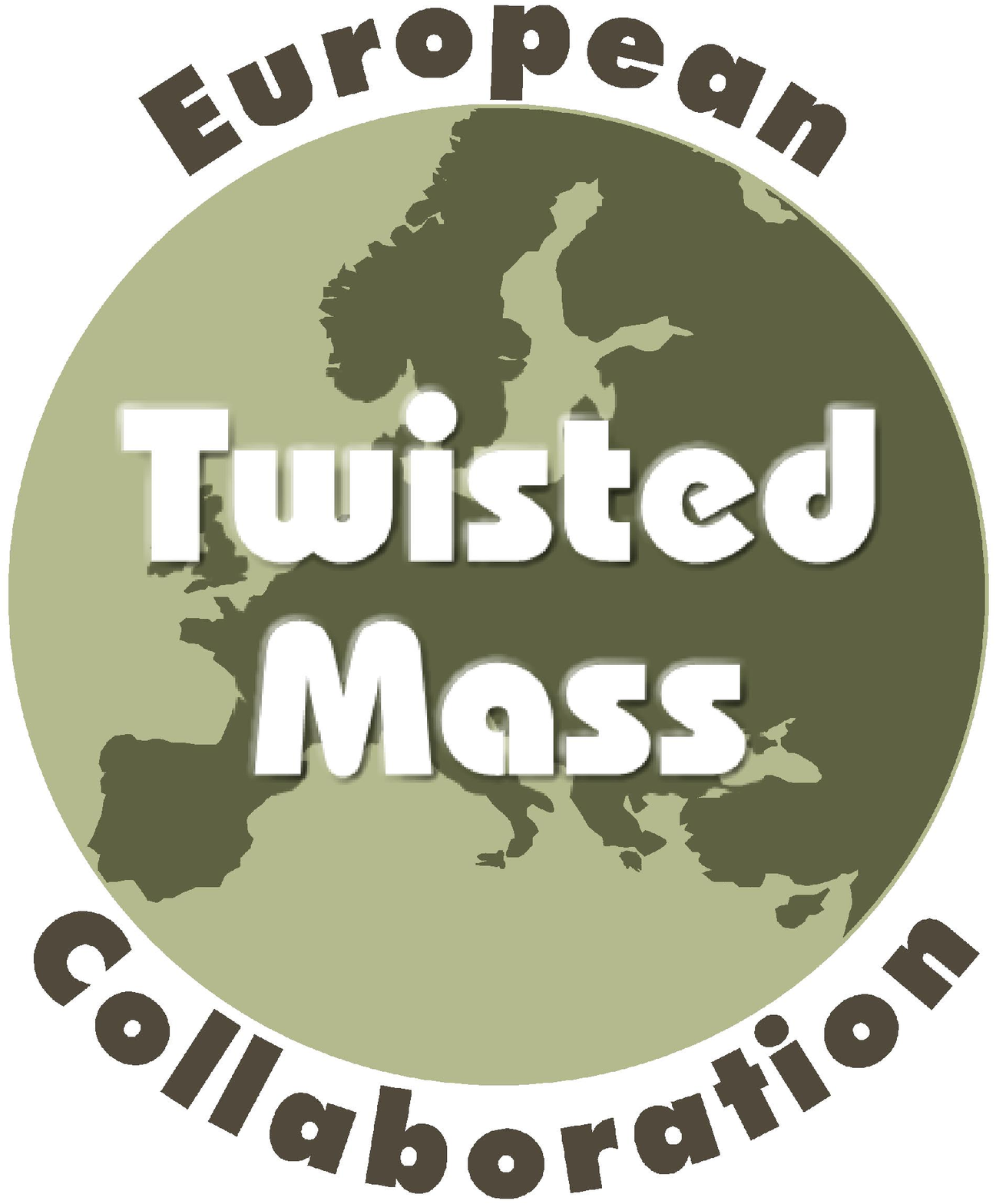}
\end{center}
}

\begin{document}
\maketitle

\section{Introduction}
One of the most important phenomena of QCD is the spontaneous breaking of chiral symmetry.
This purely non-perturbative phenomenon was subject to many analyses in Lattice QCD, using different
fermion discretizations
and methods.
In particular, the chiral condensate -- the order parameter of spontaneous chiral symmetry breaking
-- can be extracted from chiral perturbation theory fits of the quark-mass dependence of
light pseudoscalar meson observables 
\cite{Aoki:2008sm,Allton:2008pn,Aoki:2010dy,Bazavov:2009bb,Bazavov:2010yq,Baron:2009wt,Fukaya:2010na,
Bernard:2012fw,Borsanyi:2012zv}, the topological susceptibility
\cite{Durr:2001ty,Chiu:2008jq,Chiu:2011dz,Bernardoni:2010nf,Bernard:2012fw} or the pion
electromagnetic form factor \cite{Frezzotti:2008dr}.
Other methods include using the $\epsilon$-regime expansion and/or chiral random matrix theory
\cite{Hernandez:1999cu,Damgaard:2000qt,DeGrand:2005vb,Lang:2006ab,Fukaya:2007fb,Fukaya:2007yv,Fukaya:2010na,
Hasenfratz:2008ce, Jansen:2009tt, Bernardoni:2010nf,Splittorff:2012gp},
Wilson chiral perturbation theory fits of the integrated spectral density
\cite{Necco:2011vx,Necco:2011jz,Necco:2013sxa} and
a calculation directly from the quark propagator
\cite{Burger:2012ti,McNeile:2012xh}. A summary of recent results is provided in
Ref.~\cite{Colangelo:2010et}.

The chiral condensate is related to
the spectral density of the
Dirac operator via the Banks-Casher relation \cite{Banks:1979yr}:
\begin{equation}
 \label{banks-casher}
 \lim_{\lambda\rightarrow0}\lim_{m\rightarrow0}\lim_{V\rightarrow\infty}
\rho(\lambda,m)=\frac{\Sigma}{\pi},
\end{equation} 
where $\lambda$ is the modulus of the eigenvalue, $m$ the quark mass,
$\rho(\lambda,m)$ the spectral density, $V$ the volume and  $\Sigma$ is the chiral
condensate in the infinite-volume and in the chiral limit. Clearly, the triple
limit on the left-hand side of the above equation makes it impractical to
evaluate the chiral condensate on the lattice.

However, recently a method has been proposed \cite{Giusti:2008vb} to effectively make use of the
Banks-Casher relation and explore the chiral properties of QCD on the lattice,
in particular to compute the chiral condensate.
The method has also other applications, e.g. it allows to compute the topological
susceptibility or renormalization constants.
Briefly, the method consists in stochastically evaluating the {\em mode number}, 
i.e. the number of eigenmodes of
the Dirac operator below some spectral threshold value and using the dependence of this
number of eigenmodes on the threshold value to calculate the observable of interest. 
In the following, we will refer to this method as \emph{spectral projectors}.
One of its essential advantages for computing the mode number is the fact that it is
very effective in terms of computational cost -- the required computational effort grows linearly
with the lattice volume instead of quadratically, as is the case for a direct computation of
eigenmodes and counting their number below the spectral threshold. 

In this paper, we report our results for the chiral condensate
 with $N_f=2$ and $N_f=2+1+1$ Wilson twisted
mass fermions at maximal twist. Preliminary results of our computations for the $N_f=2+1+1$ case were
presented in Ref.~\cite{Cichy:2011an}. The paper is organized as follows.
In the second section we provide a short description of the spectral projector
method. In section 3, we describe our lattice setup. Section 4 presents our results for the
chiral condensate both in the $N_f=2$ and the $N_f=2+1+1$ case. In section 5, we
summarize
and compare with other determinations of the chiral condensate found in literature. An appendix
presents our tests of the method.

\section{Spectral projectors and chiral condensate}

Many interesting properties of the chiral regime of QCD can be understood from
the behaviour of quantities related to the low-lying spectrum of the Dirac
operator. One of such spectral quantities, essential in the determination of the
chiral condensate, is the {\em mode number}, i.e. the number of eigenvectors of the massive
Hermitian Dirac operator $D^\dagger D$ with eigenvalue magnitude smaller than
some threshold value $M^2$. We will denote this mode number by $\nu(M,\mu)$,
where $\mu$ is the quark mass. 

Here we provide a short description of the spectral projector method for the computation of
$\nu(M,\mu)$. For a more complete exposition, we refer to the original work of
Ref.~\cite{Giusti:2008vb}.
In this section we assume that we work on an Euclidean lattice, but
we will not specify the particular form of the lattice Dirac operator.  

If $\mathbbm{P}_M$ is
the orthogonal projector to the subspace of fermion
fields spanned by the lowest lying eigenmodes of the massive Hermitian Dirac
operator $D^\dagger D$ with eigenvalues below some threshold value $M^2$, the
mode number $\nu(M,\mu)$ can be
represented stochastically by:
\begin{equation}
 \nu(M,\mu)=\langle\textrm{Tr}\,\mathbbm{P}_M\rangle=\left\langle
\frac{1}{N}\sum_{j=1}^N (\eta_j,\mathbbm{P}_M\eta_j)\right\rangle,
\end{equation}
where $\eta_1$, $\ldots$, $\eta_N$ are pseudo-fermion fields added to the
theory.

The orthogonal projector $\mathbbm{P}_M$ can be approximated by a rational
function of $D^\dagger D$:
\begin{equation}
 \mathbbm{P}_M \approx h(\mathbbm{X})^4,\qquad
\mathbbm{X}=1-\frac{2M_*^2}{D^\dagger D+M_*^2},
\end{equation} 
where $M_*$ is a mass parameter related to the spectral threshold value $M$\footnote{As shown in
Ref.~\cite{Giusti:2008vb}, the ratio $M/M_*$ depends on the chosen approximation to the projector.
For the choice of Ref.~\cite{Giusti:2008vb}, which we apply also in our case, $M/M_*=0.96334$.}.
The function:
\begin{equation}
 h(x)=\frac{1}{2}\left(1-xP(x^2)\right)
\end{equation} 
is an approximation to the step function $\theta(-x)$ in the range $-1\leq
x\leq1$, where $P(y)$ is in our case the Chebyshev polynomial (of some
adjustable degree $n$) that minimizes the deviation:
\begin{equation}
 \delta=\max_{\epsilon\leq y\leq 1}|1-\sqrt{y}P(y)|
\end{equation} 
for some $\epsilon>0$. Computing the approximation to the
spectral projector $\mathbbm{P}_M$ requires solving the following equation an
appropriate number of times:
\begin{equation}
 (D^\dagger D + M_*^2)\psi=\eta
\end{equation} 
for a given source field $\eta$.
Solving this equation is the main computational cost in the
calculation of the mode number. In particular, the computational cost 
scales linearly $V$ with the
volume.

One can show \cite{Giusti:2008vb} that the mode number is a renormalization group
invariant, i.e.:
\begin{equation}
 \nu_R(M_R,\mu_R)=\nu(M,\mu),
\end{equation} 
where the subscript $R$ denotes renormalized quantities. Note that the spectral threshold parameter
$M$ renormalizes in the same way as the light quark mass (i.e. $M_R=Z_P^{-1}M$ for Wilson twisted
mass fermions).

Finally, we give here the relation between the mode number and the mass-dependent renormalized chiral
condensate: \cite{Giusti:2008vb} 
\begin{equation}
  \label{ch_cond}
  \Sigma_R=\frac{\pi}{2V}\sqrt{1-\left(\frac{\mu_R}{M_R}\right)^2}\frac{\partial}{\partial
  M_R}\nu_R(M_R,\mu_R),
\end{equation}
which is defined to match the chiral condensate to leading order of
chiral perturbation theory.

\section{Lattice setup}

In this section, we will specify the lattice Dirac operator that is used 
for our work, i.e. the Wilson twisted mass Dirac operator. 
For our computations of the chiral condensate, we have used gauge field configurations generated by
the European Twisted Mass Collaboration (ETMC) with 
$N_f=2$ \cite{Boucaud:2007uk,Boucaud:2008xu,Baron:2009wt} and $N_f=2+1+1$
\cite{Baron:2010bv,Baron:2010th,Baron:2011sf} dynamical quarks.

The gauge action is:
\begin{equation}
 S_G[U] = \frac{\beta}{3}\sum_x\Big( b_0 \sum_{\mu,\nu=1} \textrm{Re\,Tr} \big( 1 - P^{1\times
1}_{x;\mu,\nu}
\big) 
+ b_1 \sum_{\mu \ne \nu} \textrm{Re\,Tr}\big( 1 - P^{1 \times 2}_{x; \mu, \nu} \big) \Big),
\end{equation}
with $\beta=6/g_0^2$, $g_0$ the bare coupling and $P^{1\times
1}$, $P^{1\times 2}$ are the plaquette and rectangular Wilson loops, respectively.
For the $N_f=2$ case, we use the tree-level Symanzik improved action \cite{Weisz:1982zw}, i.e. we set
$b_1 = -\frac{1}{12}$, with the normalization condition $b_0=1-8b_1$.
In the case of $N_f=2+1+1$, we use the Iwasaki action \cite{Iwasaki:1985we,Iwasaki:1996sn}, i.e.
$b_1=-0.331$.

The Wilson twisted mass fermion action for the light, 
up and down quarks for both the $N_f=2$ and $N_f=2+1+1$ cases, is
given in the so-called twisted basis by:
\cite{Frezzotti:2000nk,Frezzotti:2003ni,Frezzotti:2004wz,Shindler:2007vp}
\begin{equation}
 S_l[\psi, \bar{\psi}, U] = a^4 \sum_x \bar{\chi}_l(x) \big( D_W + m_{0,l} + i \mu_l \gamma_5 \tau_3
\big)
\chi_l(x),
 \label{tm_light}
\end{equation}
where $m_{0,l}$ and $\mu_l$ denote, respectively, the bare untwisted and twisted light quark masses
(for shortness, whenever there is no risk of confusion, from now on we will use the symbol $\mu$ to
denote $\mu_l$).
The renormalized light quark mass is given by $\mu_R=Z_P^{-1}\mu$.
The matrix $\tau^3$ acts in flavour space and $\chi_l=(\chi_u,\,\chi_d)$ is a
two-component vector in flavour space, related to the one in the physical basis by a chiral rotation.
The standard massless Wilson-Dirac operator $D_W$ reads:
\begin{equation}
 D_W = \frac{1}{2} \big( \gamma_{\mu} (\nabla_{\mu} + \nabla^*_{\mu}) - a \nabla^*_{\mu} \nabla_{\mu}
\big),
\end{equation}
where $\nabla_{\mu}$ and $\nabla^*_{\mu}$ are the forward and backward covariant
derivatives.

The twisted mass action for the heavy doublet is given by: \cite{Frezzotti:2004wz,Frezzotti:2003xj}
\begin{equation}
 S_h[\psi, \bar{\psi}, U] = a^4 \sum_x \bar{\chi}_h(x) \big( D_W + m_{0,h} + i \mu_\sigma \gamma_5
\tau_1 + \mu_\delta \tau_3
\big)
\chi_h(x),
 \label{tm_heavy}
\end{equation}
where $m_{0,h}$ denotes the bare untwisted heavy quark mass,
$\mu_\sigma$ the bare twisted mass with the twist along the $\tau_1$ direction
and $\mu_\delta$ the mass splitting along the $\tau_3$ direction, introduced to make the strange and
charm quark masses non-degenerate.
The mass parameters $\mu_\sigma$ and $\mu_\delta$ are related to the physical renormalized strange
$m^{s}_R$ and charm $m^{c}_R$  quark masses by $m^{s,c}_R=Z_P^{-1}\left(\mu_\sigma\mp(Z_P/Z_S)\mu_\delta\right)$.
The heavy quark doublet in the twisted basis $\chi_h=(\chi_c,\,\chi_s)$ is again related to the one
in
the physical basis by a chiral rotation. 

\begin{table}[t!]
  \centering
  \begin{tabular}[]{cccccccc}
    Ensemble & $\beta$ & lattice & $a\mu$ & $\mu_R$ [MeV] &
    $\kappa_c$  & $L$ [fm]\\
\hline
  b$30.32 $ & 3.90 & $32^3\times64$ & 0.003 & 16&0.160856 & 2.7\\
   b$40.16 $  & 3.90 & $16^3\times32$   & 0.004   & 21 &0.160856   & 1.4 \\
  b$40.20 $  & 3.90 & $20^3\times40$  & 0.004 &  21&  0.160856   &  1.7\\
  b$40.24 $  & 3.90 & $24^3\times48$  & 0.004 &  21&  0.160856   &  2.0 \\
  b$40.32$ &  3.90 & $32^3\times64$  & 0.004 & 21& 0.160856 & 2.7 \\
 b$64.24$ & 3.90 & $24^3\times48$  & 0.0064 &34 &0.160856 & 2.0\\
 b$85.24$ & 3.90 & $24^3\times48$  & 0.0085 & 45& 0.160856 & 2.0 \\
  c$30.32$ & 4.05 & $32^3\times64$  & 0.003 & 19 &0.157010 &  2.1\\
  c$60.32$ & 4.05 & $32^3\times64$  & 0.006 & 37 &0.157010 &  2.1\\
  c$80.32$ & 4.05 & $32^3\times64$  & 0.008 & 49 &0.157010 & 2.1\\
  d$20.48$ & 4.20 & $48^3\times 96$  & 0.002 & 15 &0.154073 &  2.6\\
  d$65.32$ & 4.20 & $32^3\times64$  & 0.0065 & 47 &0.154073 & 1.7\\
  \end{tabular}
  \caption{Parameters of the $N_f=2$ gauge ensembles
\cite{Boucaud:2007uk,Boucaud:2008xu,Baron:2009wt}. We show the inverse bare coupling $\beta$,
lattice size $(L/a)^3\times(T/a)$, bare twisted light quark mass $a\mu$,
renormalized quark mass $\mu_R$ in MeV, critical value of the hopping parameter at which the
PCAC mass vanishes and physical extent of the lattice $L$ in fm.}
  \label{setupNf2}
\end{table}

The twisted mass formulation allows for an automatic $\mathcal{O}(a)$
improvement of physical observables, provided the hopping parameter $\kappa = (8+2 a m_0)^{-1}$,
where $m_0\equiv m_{0,l}=m_{0,h}$ can be chosen, is
tuned to maximal twist by setting it to its critical value, at which the PCAC quark mass vanishes
\cite{Frezzotti:2003ni,Chiarappa:2006ae,Farchioni:2004ma,Farchioni:2004fs,Frezzotti:2005gi,
Jansen:2005kk}.

\begin{table}[t!]
   \centering
  \begin{tabular}[]{cccccccc}
    Ensemble & $\beta$ &       lattice & $a\mu_l$ & $\mu_{l,R}$ [MeV]&
    $\kappa_c$  & L  [fm]\\
\hline
A30.32 &1.90 & $32^3\times 64$  & 0.0030 & 13 & 0.163272  & 2.8 \\
A40.20 &1.90 & $20^3\times 40$  & 0.0040 & 17 & 0.163270  & 1.7 \\
A40.24 &1.90 & $24^3\times 48$  & 0.0040 & 17 & 0.163270  & 2.1 \\
A40.32 &1.90 & $32^3\times 64$  & 0.0040 & 17 & 0.163270  & 2.8 \\
A50.32 &1.90 & $32^3\times 64$  & 0.0050 & 22 & 0.163267  & 2.8 \\
A60.24 &1.90 & $24^3\times 48$  & 0.0060 & 26 & 0.163265  & 2.1 \\
A80.24 & 1.90 & $24^3\times 48$  & 0.0080 & 35 & 0.163260  & 2.1 \\
 B25.32 & 1.95 & $32^3\times64$  & 0.0025 & 13& 0.161240 & 2.5\\
  B35.32 &1.95 & $32^3\times64$  & 0.0035 & 18& 0.161240 &2.5\\
  B55.32 & 1.95 & $32^3\times64$  & 0.0055 &28& 0.161236 &2.5\\
  B75.32 &1.95 & $32^3\times64$  & 0.0075 &38 &0.161232 &2.5\\
  B85.24 & 1.95 & $24^3\times48$  & 0.0085  &45 &0.161231 &1.9 \\
  D15.48 & 2.10 & $48^3\times96$  &  0.0015 & 9 &0.156361 & 2.9\\
  D20.48 &  2.10 & $48^3\times96$  & 0.0020 & 12 &0.156357&2.9\\
  D30.48 & 2.10 & $48^3\times96$  & 0.0030 & 19 &0.156355 &2.9\\
   \end{tabular}
  \caption{Parameters of the $N_f=2+1+1$ gauge ensembles
\cite{Baron:2010bv,Baron:2010th,Baron:2011sf}. We show the inverse bare coupling $\beta$,
lattice size $(L/a)^3\times(T/a)$, bare twisted light quark mass $\mu_l$,
renormalized quark mass $\mu_{l,R}$ in MeV, critical value of the hopping parameter at which the
PCAC mass vanishes and physical extent of the lattice $L$ in fm.}
  \label{setupNf211}
\end{table}

\begin{table}[t!]
  \centering
  \begin{tabular}[]{ccccc}
    $N_f$ & $\beta$ & $a$ [fm] & $Z_P(\MSb,\,2\,{\rm GeV})$ & $r_0/a$\\
\hline
2 & 3.90 & 0.085 & 0.437(7) & 5.35(4)\\
2 & 4.05 & 0.067 & 0.477(6) & 6.71(4)\\
2 & 4.20 & 0.054 & 0.501(13) & 8.36(6)\\
2+1+1 & 1.90 & 0.0863 & 0.529(9) & 5.231(38)\\
2+1+1 & 1.95 & 0.0779 & 0.504(5) & 5.710(41)\\
2+1+1 & 2.10 & 0.0607 & 0.514(3) & 7.538(58)\\
\end{tabular}
  \caption{The values of the lattice spacing $a$ \cite{Blossier:2010cr,Baron:2011sf}, $r_0/a$
\cite{Blossier:2010cr,Baron:2010bv,Ottnad:2012fv} and the renormalization constant $Z_P$ in
the $\MSb$ scheme at
the scale of 2 GeV \cite{Constantinou:2010gr,Alexandrou:2012mt,Palao:priv}, for different
values of $\beta$ and $N_f=2$ and $N_f=2+1+1$.}
\label{tab:ZP}
\end{table}

The details of the ensembles considered for this work are presented in Tab. \ref{setupNf2} for
$N_f=2$ and Tab. \ref{setupNf211} for $N_f=2+1+1$.
For both cases, they include 3 lattice spacings (from $a\approx0.05$ to $a\approx0.085$ fm) and up to
5 quark masses at a given lattice spacing.
The renormalized light quark masses $\mu_R$ are in the range from around 15 to 50 MeV.
The values of the renormalization constant $Z_P$ for different ensembles\footnote{For
$N_f=2+1+1$, the mass-independent renormalization constant $Z_P$ is extracted as a chiral limit of a
dedicated computation with 4 mass-degenerate flavours -- see Refs.~\cite{Dimopoulos:2011wz,ETM:2011aa}
for details.}
\cite{Constantinou:2010gr,Alexandrou:2012mt,Palao:priv}, used to convert
bare light quark masses $\mu$
and bare spectral threshold parameters $M$ to their renormalized values in the $\MSb$ scheme (at the
scale of 2 GeV), are given in Tab.~\ref{tab:ZP}, where we also show the values of $r_0/a$ (in the
chiral limit), used to
express our results for the condensate as a dimensionless product $r_0\Sigma^{1/3}$.
Our physical lattice extents $L$ for extracting physical results range from 2 to 3 fm (in the
temporal direction, we always have $T=2L$).
To check for the size of finite volume effects, we included different lattice sizes for $\beta=3.9$,
$a\mu=0.004$ ($N_f=2$) and $\beta=1.9$, $a\mu=0.004$ ($N_f=2+1+1$).

\section{Results}

In this section, we show our results of the 
calculation of the chiral condensate.
First, we illustrate the procedure of extraction of the chiral condensate and discuss the influence
of the various errors that enter the computation.
Then, we analyze finite volume effects in our simulations.
Finally, we move on to our chiral and continuum extrapolations.

\subsection{Procedure and errors}
\label{sec:procedure}
We show here how to extract the mass-dependent chiral condensate according to Eq.~\eqref{ch_cond},
illustrating the procedure for ensemble B40.32. Using the spectral projector method, we computed the
dependence of the mode number on the renormalized spectral threshold parameter $M_R$ for 5 values of
$M_R$, from around 2.5 times the renormalized quark mass to around 120 MeV. Shortly above the
latter value one starts to see deviations from the linear regime of $\nu_R(M_R,\mu_R)$ vs. $M_R$ (see
Appendix \ref{app:tests}).

Fig.~\ref{procedure} shows the dependence of the mode number on the renormalized spectral
threshold parameter $M_R$. The solid line is a linear fit to all 5 points.
The slope of this line $\partial\nu_R(M_R,\mu_R)/\partial M_R$ determines the value of the
mass-dependent chiral condensate according to Eq.~\eqref{ch_cond}.
The error of this slope includes two sources: the error of the slope of
the bare mode number as function of the bare threshold parameter $M$, 
$\partial\nu(M,\mu)/\partial M$ and the error of $Z_P$ needed to convert from bare to renormalized
quantities. Although $\partial\nu_R(M_R,\mu_R)/\partial M_R$ appears 
to be constant as a function of $M_R$ within errors,  we will take
its value to be the middle point of the chosen fitting interval, see below for details 
of the fitting intervals considered.
Finally, Eq.~\eqref{ch_cond} yields, after taking the cubic root\footnote{The values of
$\Sigma$ that we give (also in our plots) are always for the renormalized condensate.}:
\begin{displaymath}
 a\Sigma^{1/3}=0.13372(34)(72),
\end{displaymath}
where the first error is the one of the slope $\partial\nu(M,\mu)/\partial M$ and the other one comes
from $Z_P=0.437(7)$ and is dominated by systematic effects -- hence, we take it as a systematic error
of our computation.

\begin{figure}[t!]
  \begin{center}
    \includegraphics[width=0.6\textwidth,angle=270]{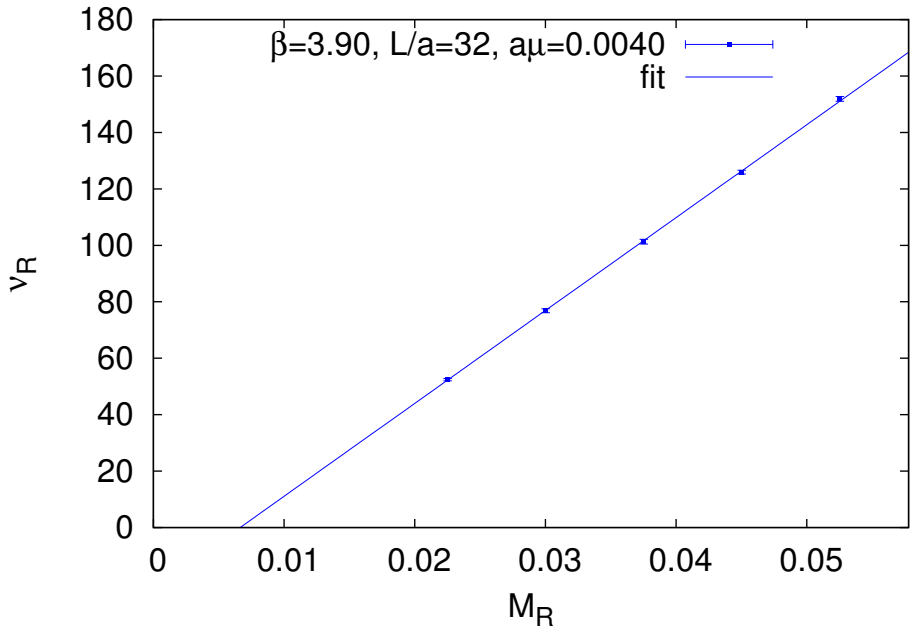}
  \end{center}
  \caption{\label{procedure} Dependence of the mode number on the renormalized spectral
threshold parameter $M_R$ for ensemble B40.32. The solid line is a linear fit to all 5 points.} 
\end{figure}

\begin{table}[t!]
  \centering
  \begin{tabular}[]{ccccc}
    fit range & lowest $aM_R$ & highest $aM_R$ & $\chi^2/{\rm dof}$ & $r_0\Sigma^{1/3}$\\
\hline
1 -- 3 & 0.0225 & 0.0375 & 0.004 & 0.7085(43)\\
1 -- 4 & 0.0225 & 0.0450 & 0.018 & 0.7116(24)\\
1 -- 5 & 0.0225 & 0.0525 & 0.588 & 0.7154(18)\\
2 -- 4 & 0.0300 & 0.0450 & 0.006 & 0.7141(46)\\
2 -- 5 & 0.0300 & 0.0525 & 0.567 & 0.7189(33)\\
3 -- 5 & 0.0375 & 0.0525 & 0.549 & 0.7235(57)\\
\end{tabular}
\caption{Values of $r_0\Sigma^{1/3}$ extracted from different fitting ranges. Every fit includes
at least 3 values of $M_R$. The fit labeled ``1 -- 5'' is the full fit. We estimate the error from
the choice of the fitting range by comparing the value from the full fit with the ones from fits ``1
-- 4'' and ``2 -- 5''. The error given is statistical only.}
\label{tab:fitrange}
\end{table}

The value of $a\Sigma^{1/3}$ can be further converted to a dimensionless product $r_0\Sigma^{1/3}$
(which will be the final result of this paper, after taking the chiral and continuum limits) or to a
physical value in MeV. For the former, we use the value in the chiral limit $r_0/a=5.35(4)$. Since the
error of this value is again mostly systematic, we quote it as another systematic error of
$r_0\Sigma^{1/3}$:
\begin{displaymath}
 r_0\Sigma^{1/3}=0.7154(18)(39)(53),
\end{displaymath}
where the two errors are as above and the third one comes from $r_0/a$.
For a conversion to MeV, one needs to choose a value of the lattice spacing in physical units.
There are several such estimates for ETMC 2-flavour ensembles, giving for $\beta=3.9$ values
including 0.079 fm \cite{Baron:2009wt}, 0.085 fm \cite{Blossier:2010cr} and
0.089 fm \cite{Alexandrou:2008tn}. 
Taking this spread into account, the relative error on the lattice spacing is around 7\%, which
leads to a similar relative uncertainty in the value of the chiral condensate $\Sigma^{1/3}$ in
physical units, which amounts to about 20 MeV.
This is roughly an order of magnitude more than other errors entering our computation.
Even being less conservative and using for the
error the value quoted in Ref.~\cite{Blossier:2010cr} -- 0.085(2)$_{\rm stat}$(1)$_{\rm syst}$ fm --
the
error that it yields is still of the order of 10 MeV. Therefore, we decided to give our final results
as the dimensionless product $r_0\Sigma^{1/3}$ and we chose not to quote any value in MeV for it until
a significantly improved determination of the lattice spacing 
is available.

Another source of the error is the choice of the fitting range and hence the value of $M_R$ that
enters the square root in Eq.~\eqref{ch_cond}. Of course, physical results should not depend on
this choice, provided that the whole fitting interval lies in the linear regime of the mode
number vs. $M_R$ dependence.
Hence, varying the fitting range serves two purposes: establishing whether non-linear effects are
already present and checking that the choice of $M_R$ in Eq.~\eqref{ch_cond} does not influence the
final result.
The values of $r_0\Sigma^{1/3}$
resulting from different fitting ranges in $M_R$ are shown in Tab.~\ref{tab:fitrange}. For all fits,
$\chi^2/{\rm d.o.f.}$ is below 1. The compatibility of all results and the good values of
$\chi^2/{\rm d.o.f.}$ imply that for this ensemble the choice of the fitting range and $M_R$ does not
affect the results in a substantial way\footnote{The ensemble B40.32 is somewhat special in this
aspect. As we show below, in general, the fitting range uncertainty is the most important source of
error in our analysis.}.
To quantify this error, we considered the 4-point fits with the lowest
or highest value of $M_R$ excluded.
Excluding the first or last point leads to a similar change of the result and hence we took the
larger of the two as our conservative error from the choice of the fitting range.
Note, however, that Tab.~\ref{tab:fitrange} implies a systematic tendency towards
increasing of
$\Sigma$ when the fitting range moves towards higher values of $M_R$. This indicates an onset of
non-linear behaviour for values of $M_R$ only slightly above the ones we considered.

Finally, our estimate of $r_0\Sigma^{1/3}$ for ensemble B40.32, including all sources of error, is:
\begin{displaymath}
 r_0\Sigma^{1/3}=0.7154(18)_{\rm stat}(38)_{\rm fit}\,_{\rm range}(39)_{Z_P}(53)_{r_0/a}.
\end{displaymath} 
We note that the total error is dominated by systematic errors. This means that increasing statistics
would not essentially change our total error. It should be considered an important advantage of the
method of spectral projectors that rather moderate statistics (in our case around 230 independent
gauge field configurations for this ensemble) leads to a practically negligible statistical error.
Let us also mention that the quoted statistical error takes autocorrelations fully into
account. We performed an analysis of autocorrelations using two methods and found that in general the
autocorrelations are small, even at our smallest lattice spacings. For the details of our
autocorrelation analysis, we refer to Appendix \ref{sec:tauint}.

\subsection{Finite volume effects}
One of the main sources of systematic effects in Lattice QCD simulations are finite volume effects
(FVE).
In Ref.~\cite{Giusti:2008vb}, theoretical arguments were provided that FVE should be small for
the chiral condensate computed from the mode number -- with exponentially small difference between the
finite volume and infinite volume results of $\mathcal{O}(\exp(-M_\Lambda L/2))$, where
$M_\Lambda^2=2\Lambda\Sigma/F^2$, $\Lambda=\sqrt{M^2-\mu^2}$ and $F$ is the pion decay constant in the
chiral limit. Since in practice the mass-dependent chiral condensate is extracted at
$\Lambda\gg\mu$,
the mass $M_\Lambda$ is much higher than the pion mass, which typically governs FVE.
Hence, one expects that for the computation of the chiral condensate from the mode number, FVE will be
rather small.
FVE for the mode number itself were computed in SU(2) chiral perturbation theory
\cite{Necco:2011vx}.
The resulting formula leads to a prediction that FVE from lattices with $L\geq2$ fm should be small,
$\mathcal{O}(\lesssim1\%)$ for $M_R\approx\mathcal{O}(60-120)$ MeV and renormalized quark masses of
$\mathcal{O}(10-20)$ MeV (with larger FVE at smaller $M_R$). Indeed, in practice, it was shown
in Ref.~\cite{Giusti:2008vb} that for $L\geq2$
fm the results deviate from their infinite volume values by less than 1\%.

To show that it is also the case in our setup, we performed the computation of the mode number and
the chiral condensate for:
\begin{itemize}
 \item $N_f=2:\;$ 4 different volumes at fixed $\beta=3.9$, $a\mu=0.004$, lattice extents:
$L/a=16$, 20, 24 and 32, with corresponding physical values of 1.4, 1.7, 2.0 and 2.7 fm, respectively,
 \item $N_f=2+1+1:\;$ 3 different volumes at fixed $\beta=1.9$, $a\mu=0.004$, lattice
extents: $L/a=20$, 24 and 32, with corresponding physical values of 1.7, 2.1 and 2.8 fm,
respectively.
\end{itemize}

\begin{figure}[t!]
\begin{minipage}[b]{0.5\linewidth}
\includegraphics[width=0.74\textwidth,angle=270]{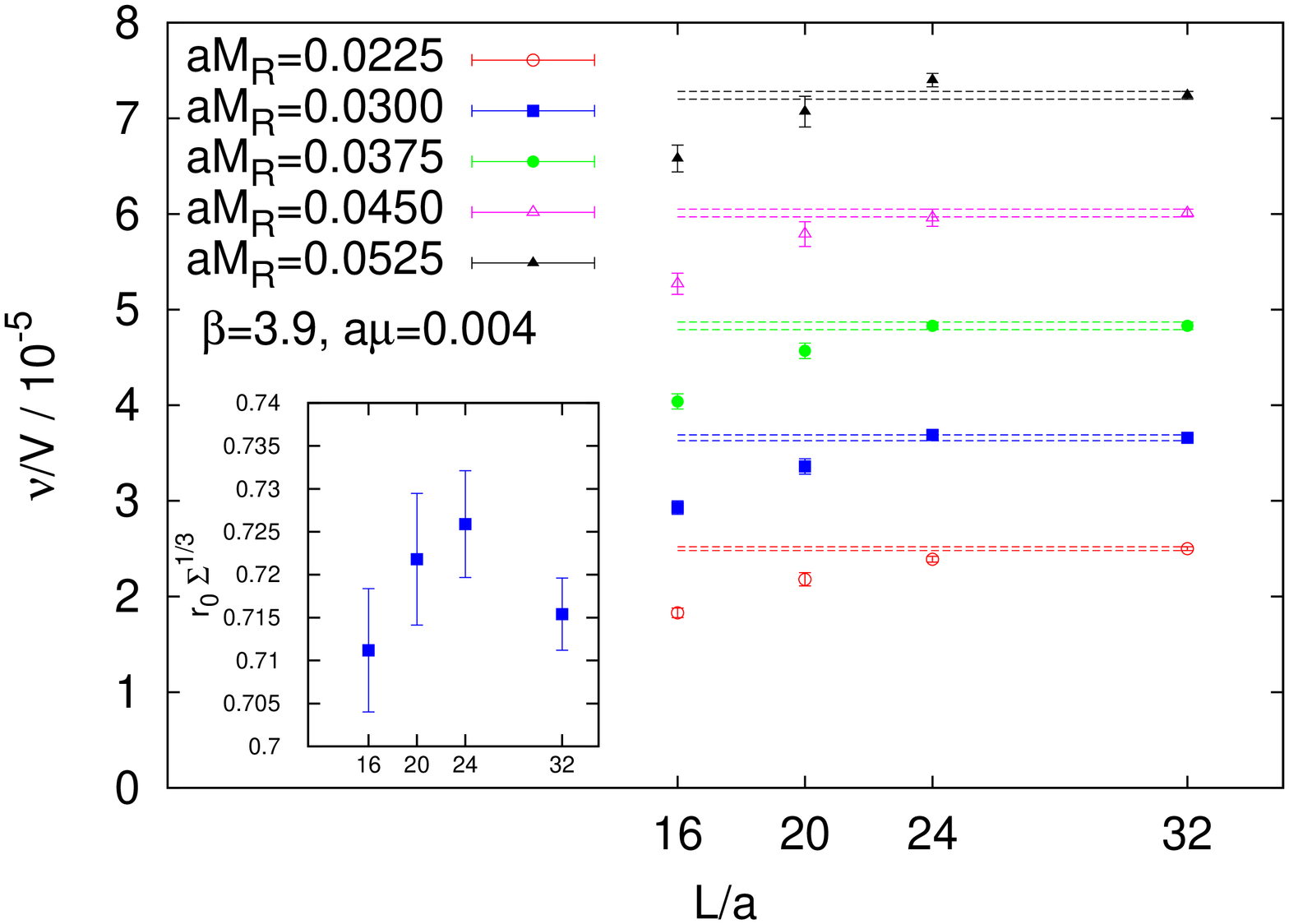}    
\label{fve2}
\end{minipage}
\begin{minipage}[b]{0.5\linewidth}
\includegraphics[width=0.74\textwidth,angle=270]{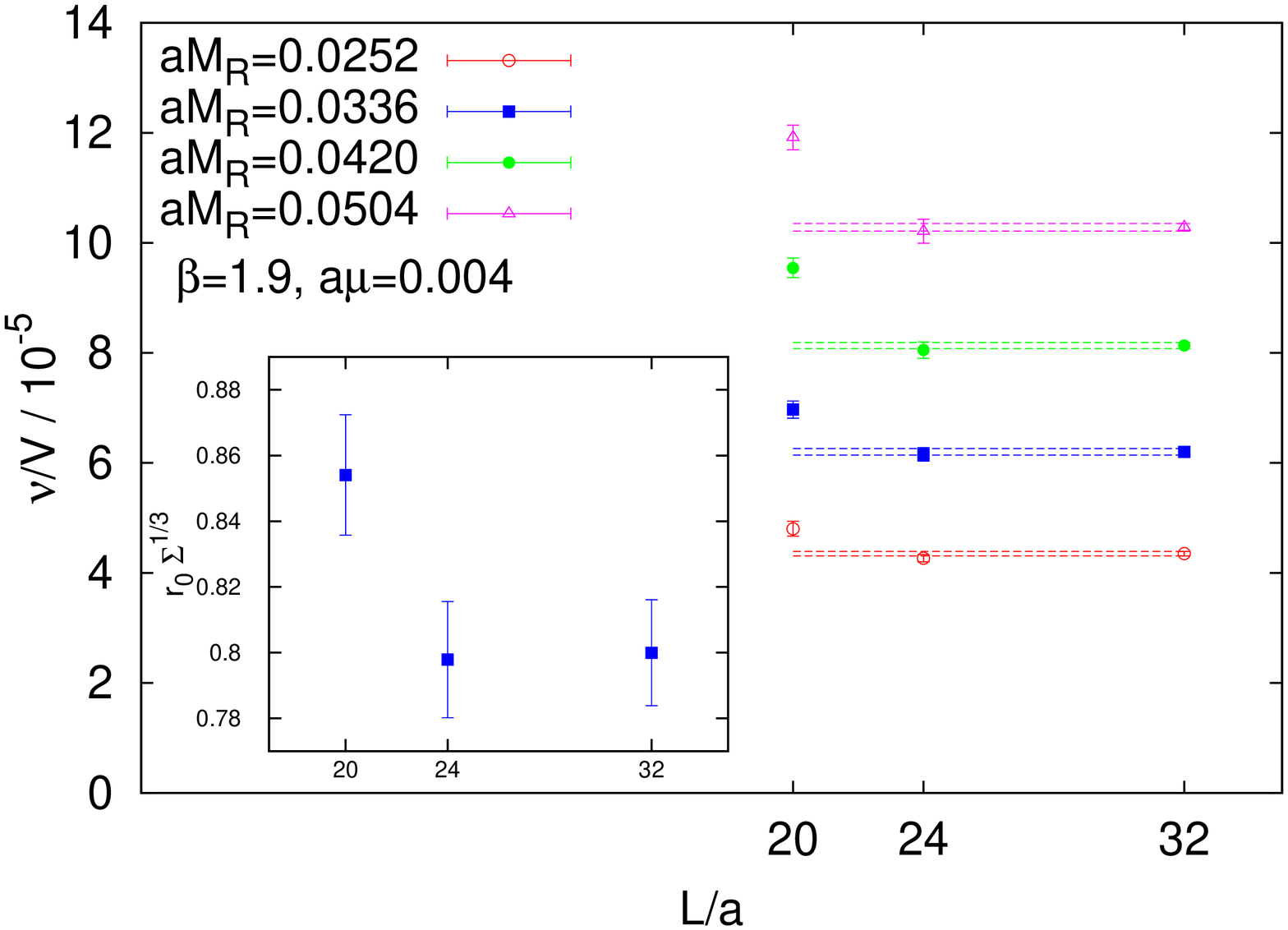}
\label{fve211}
\end{minipage}
\caption{The main plots show the volume dependence of the mode number density $\nu/V$ for different
values of the renormalized spectral threshold $M_R$. The horizontal bands show the result
at the largest volume. The insets show the
volume dependence of the chiral condensate $r_0\Sigma^{1/3}$ (the error of each point includes the
statistical error and the systematic one originating from the choice of the fitting interval).
(Left) $N_f=2$, $\beta=3.9$, $a\mu=0.004$, (right) $N_f=2+1+1$, $\beta=1.9$, $a\mu=0.004$.}
\label{fve_inset}
\end{figure}

In Fig.~\ref{fve_inset}, we show the volume dependence of the mode number density
$\nu/V$ for 4-5 different values of the renormalized spectral threshold $M_R$. 
The mode number density can be computed very precisely.
It hence provides a strong test of finite size effects.

The left plot shows our data for $N_f=2$.
The results for $L/a=20$ and especially $L/a=16$ are systematically lower than $L/a=32$, signaling
large FVE.
However, the mode number density for $L/a=24$ is compatible with the one for $L/a=32$ for 3
intermediate values of $M_R$, while it differs by 2-3$\sigma$ for the lowest and highest $M_R$,
thus changing the slope of the mode number vs. $M_R$ dependence and the extracted chiral condensate
(see the inset of Fig.~\ref{fve_inset} (left)). This change of slope is statistically significant, but
it is still a relatively small, 1-1.5\% effect.
Taking into account the uncertainty from the choice of the fitting range, the final results for the
chiral condensate are compatible for all cases -- including the ones for small volumes, indicating
that even if the mode number density goes
systematically down, the slope of the whole $\nu(M_R,\mu_R)$ dependence is less affected.
We have also tried a description of FVE in the framework of the formula derived in
Ref.~\cite{Necco:2011vx}. We conclude that it provides a reasonable agreement with actual lattice
data for $L/a\gtrsim24$, while FVE for smaller volumes are somewhat underestimated
(by a factor of $\mathcal{O}(2)$ at $L/a=16$, compared to the actually observed FVE).

In the right plot, we show analogous data for $N_f=2+1+1$. Similarly, we observe significant finite
size effects in the mode number density (and also in the chiral condensate) for $L/a=20$, while
$L/a=24$ and $L/a=32$ are always compatible.

This allows us to conclude that finite size effects are indeed very small when one
reaches a linear lattice extent of around 2 fm ($L/a=24$ at both $\beta=3.9$ ($N_f=2$) and
$\beta=1.9$ ($N_f=2+1+1$)).
Therefore, we used in our analysis all available ETMC ensembles with a linear lattice extent of at
least 2 fm\footnote{The exception to this rule is ensemble d65.32 with $L\approx1.7$ fm. We decided
to use this ensemble, because without it there would only be one quark mass at $\beta=4.2$ and no
chiral extrapolation could be performed.}.

\subsection{Chiral and Continuum Limit -- $N_f=2$}

We now show our results for the 2-flavour case.
For each value of $\beta$, we have 2-4 sea quark masses, according to Tab.~\ref{setupNf2}. For each
ensemble, we perform computations of the mode number at 5 values of the renormalized spectral
threshold $M_R$, from around 50 to 120 MeV. We follow the procedure outlined in
Sec.~\ref{sec:procedure}, i.e. we extract the mass-dependent condensate from the slope 
$\nu_R(M_R,\mu_R)$ as function of $M_R$ for each ensemble.

\begin{table}[t!]
  \centering
  \begin{tabular}[]{ccc}
    Ensemble & $a\mu$ & $r_0\Sigma^{1/3}$\\
\hline
  b$30.32 $ & 0.0030 & 0.7118(29)(38)(53)\\
  b$40.32$ & 0.0040 & 0.7154(18)(39)(53)\\
 b$64.24$ & 0.0064 & 0.7246(26)(39)(54)\\
 b$85.24$ & 0.0085 & 0.7377(23)(40)(55)\\
 chiral & & 0.6957(35)(37)(52)(186)\\
\hline
  c$30.32$ & 0.0030 & 0.7188(50)(30)(43)\\
  c$60.32$ & 0.0060 & 0.7345(35)(30)(44)\\
  c$80.32$ & 0.0080 & 0.7425(29)(31)(44)\\
chiral & & 0.7046(78)(30)(42)(206)\\
\hline
  d$20.48$ & 0.0020 & 0.7036(45)(61)(51)\\
  d$65.32$ & 0.0065 & 0.7415(40)(64)(53)\\
chiral & & 0.6853(73)(59)(49)(265)\\
  \end{tabular}
  \caption{Results for $r_0\Sigma^{1/3}$ for all considered $N_f=2$ ensembles. The given errors are:
statistical, resulting from $Z_P$, resulting from $r_0/a$, respectively. We also show results in the
chiral limit, where we also give the systematic error from the choice of the fitting interval (4th
error). See text for more details.}
  \label{tab:results_Nf2}
\end{table}

\begin{figure}[t!]
  \centering
  \includegraphics[width=0.7\textwidth,angle=270]{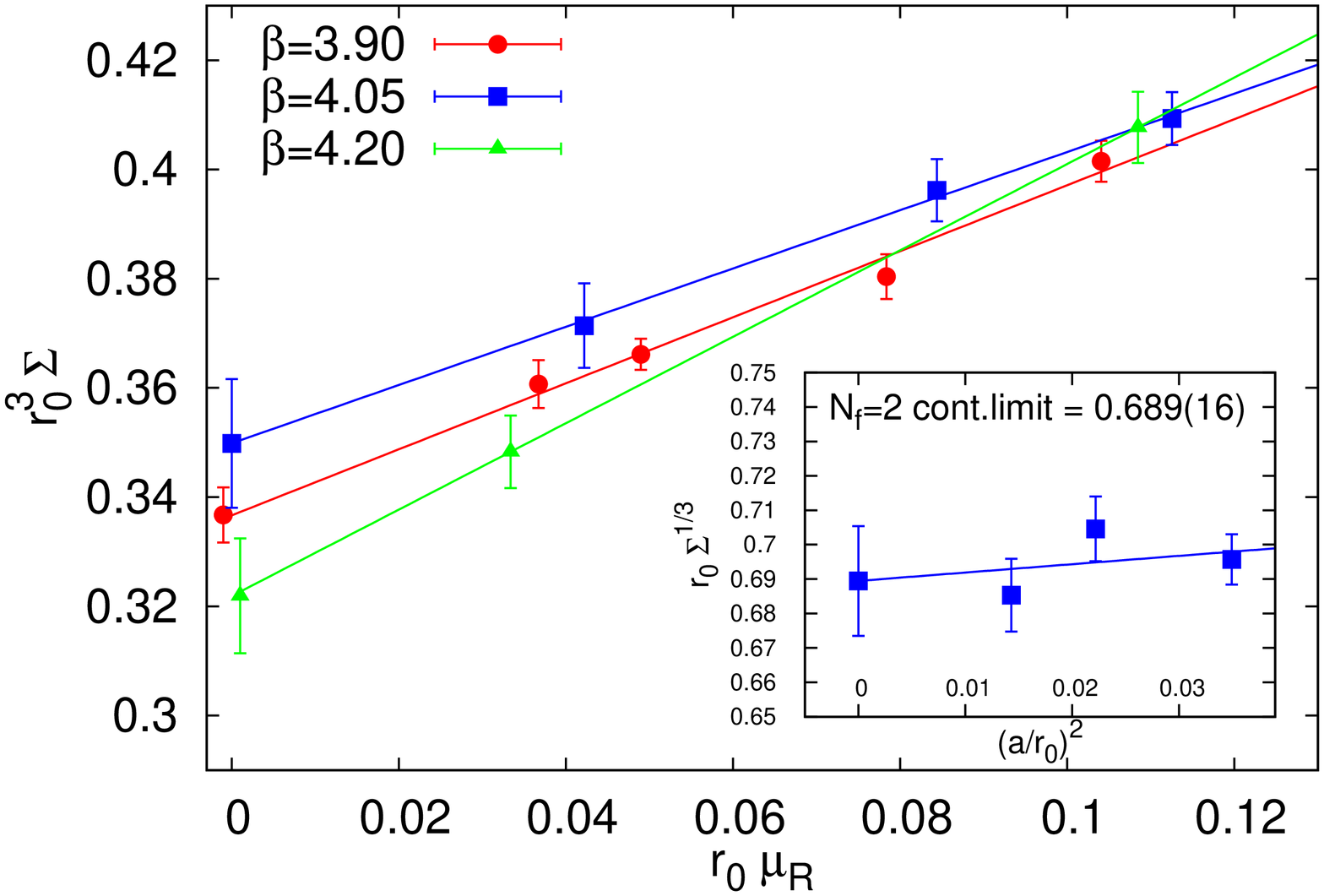}
  \caption{Main plot: chiral extrapolations of the chiral condensate $r_0^3\Sigma$ for 
    $N_f=2$ ensembles and $\beta=3.9$, 4.05, 4.2. The lines are extrapolations to the
chiral limit, linear in the quark mass. The values in the chiral limit for $\beta=3.9$ and 4.2 are
slightly shifted for better presentation. The errors are statistical only. Inset: continuum
extrapolation of the chirally extrapolated chiral condensate $r_0\Sigma^{1/3}$ vs. $(a/r_0)^2$. The
errors include: statistical errors, errors from $Z_P$ and errors from $r_0/a$.}
 \label{fig:chiral_Nf2}
\end{figure}

The results for $r_0\Sigma^{1/3}$ for all considered ensembles are gathered in
Tab.~\ref{tab:results_Nf2}.
These results are then used to extrapolate to the chiral limit for each value of $\beta$.
The chiral corrections to the mass-dependent condensate were calculated at the
next-to-leading order of chiral perturbation theory in Ref.~\cite{Giusti:2008vb}.
The obtained formula suggests that the mass-dependent condensate is equal to
the chiral condensate in the chiral limit up to terms linear in $\mu_R$ and higher order effects. In
particular, there are no corrections proportional to $\mu_R\,\ln\mu_R$.
Moreover, the size of these chiral corrections is small, as illustrated explicitly in
Ref.~\cite{Giusti:2008vb} -- inserting the values of low energy constants, it was shown that
regardless of the value of $M_R$ at which the condensate is extracted, the curvature is very mild.
Hence, in practice a linear extrapolation of the mass-dependent condensate to the chiral limit is
fully justified and we follow this conclusion in our chiral extrapolations.
As a check, we tried fits of the NLO formula (inserting values of the low energy constants used in
Ref.~\cite{Giusti:2008vb}) and we found that the differences with respect to the linear extrapolation
are negligible compared to our errors.

Our extrapolations for all three values of $\beta$ are shown in the main plot of
Fig.~\ref{fig:chiral_Nf2} (we plot $r_0^3\Sigma$ vs. $r_0\mu_R$ to allow comparisons between
different
values of $\beta$).
The plotted errors are only statistical, since in extrapolations at fixed $\beta$, the relative errors
from $Z_P$ and $r_0/a$ are the same for all quark mass values (we use chirally extrapolated values of
$Z_P$ and $r_0/a$) -- we give them in Tab.~\ref{tab:results_Nf2}.
To estimate the systematic error originating from the choice of the fitting range in
$\nu_R(M_R,\mu_R)$ vs. $M_R$ fits, we repeated all chiral extrapolations for two tailored fitting
ranges -- excluding the first value of $M_R$ (to account for effects of coming too close to the
renormalized quark mass) or the last value thereof (to account for possible deviations from the
linear behaviour for too high values of $M_R$).

The chiral limit values, with all sources of error, are also shown in Tab.~\ref{tab:results_Nf2}.
In general, the total error originates in practice only from the choice of the fitting
range and the latter increases when approaching the continuum limit. The reasons for this behaviour
include the fact that the number of quark masses that we use
decreases for smaller lattice spacings and at $\beta=4.2$ the slope of the
quark mass dependence of the chiral condensate is apparently larger than at coarser lattice
spacings\footnote{Note that this slope may be affected by the smaller volume of ensemble d65.32 --
hence, it may be a residual FVE and not an indication of the dependence of the
slope on the lattice spacing. In such case, our fitting range error at $\beta=4.2$ implicitly
reflects this FVE.}, making the final chiral limit value more susceptible to changes in
the fitting interval.

Finally, we can use the chirally extrapolated values of the condensate to perform an extrapolation to
the continuum limit.
We start by discussing the $\mathcal{O}(a)$-improvement of the chiral condensate.
For on-shell quantities, $\mathcal{O}(a)$-improvement amounts to the quantity being even 
under the $\mathcal{R}_5$ parity transformation:
%$\psi\rightarrow \gamma_5\psi$, $\bar \psi
%\rightarrow -\bar\psi \gamma_5$ \cite{Frezzotti:2003ni}.
$\psi\rightarrow i\gamma_5\tau^1\psi$, $\bar \psi
\rightarrow i\bar\psi \gamma_5\tau^1$ \cite{Frezzotti:2003ni}.
Let us consider the spectral sums \cite{Giusti:2008vb}:
$\sigma_k(\mu, m_q)=\langle \mathrm{Tr}\left\{ (D_m^{\dagger}D_m +\mu^2)^{-k}\right\} \rangle$,
where $k\geq3$ for reasons explained in the given reference.
The spectral sums are related to the
mode number \cite{Giusti:2008vb} and the improvement (or lack thereof) of the spectral sums implies
the improvement of the chiral condensate. Representing
the spectral sums as a density chain correlation function (for $k=3$):
\begin{equation}
  \label{eq:14}
  \sigma_3(\mu,m)=-a^{24} \sum_{x_1...x_{6}}\langle P_{12}^+(x_1) P_{23}^-(x_2) P^+_{34}(x_{3})
P^-_{45}(x_{4}) P^+_{56}(x_{5}) P^-_{61}(x_{6}) \rangle,
\end{equation}
it is straightforward to show that the object on the right-hand side is
even under $\mathcal{R}_5$ transformation, since the number of densities is even.
However, one also needs to consider contact terms arising from Eq.~\eqref{eq:14}, i.e. terms in the
sum with $x_i=x_j$ for some $i\neq j$.
It can be demonstrated \cite{Oa-impr} that such terms give rise only to $\mathcal{O}(am_0)$ terms in
the mode number -- hence they vanish at maximal twist.
In this way, the contact terms do not spoil automatic $\mathcal{O}(a)$-improvement of the chiral
condensate.

Hence, our continuum limit extrapolation is performed linearly in $a^2$, using results at three
lattice spacings, with fixed fitting range of the mode number vs. spectral threshold dependence,
corresponding
to $M_R\approx90$ MeV (entering the square root in Eq.~\eqref{ch_cond}) for all values of $\beta$. As
an
error, we use the statistical error, combined in quadrature with the error of $Z_P$ and $r_0/a$. We do
not observe significant cut-off effects. The final value in the continuum limit is 0.689(16).
To account for the fitting range error, we perform the full analysis for tailored fitting ranges,
excluding the first or last value of $M_R$ for
each ensemble.
This corresponds to a shift in $M_R$ to approx. 80 or 100 MeV, respectively.
While the extracted value of the condensate in the chiral limit should not depend on the fitting
range, in practice the results for different fitting ranges differ, which is due to using only 4-5
values in the fits to extract the slope of $\nu_R(M_R,\mu_R)$. The fits from tailored fitting ranges
yield 0.678(18) and 0.718(20), respectively.
To be conservative, as our systematic error from the fitting range we choose the larger difference of
the two with respect to the central value 0.689(16).
This finally gives:
\begin{displaymath}
 r_0\Sigma^{1/3}_{N_f=2}=0.689(16)(29),
\end{displaymath}
where the first error is the combined statistical error, the error of $Z_P$ and of $r_0/a$, while the
second error originates from the choice of the fitting range.

\subsection{Chiral and Continuum Limit -- $N_f=2+1+1$}
In this subsection, we present results for the $N_f=2+1+1$ case. By comparing to the results of the
2-flavour case, we can investigate the role of the dynamical strange and charm quarks.

We proceed as in the previous section. For each value of $\beta$, we have 3-5 sea quark masses,
according to Tab.~\ref{setupNf211}. We compute the mode number at 4 values of
the renormalized spectral threshold $M_R$, from around 50 to 110 MeV, and extract the mass-dependent
condensate from the slope of the $\nu_R(M_R,\mu_R)$ vs. $M_R$ dependence for each ensemble.
We have also computed the mode number for a fifth value of $M_R\approx130$ MeV.
However, given the very good statistical precision of the spectral 
projectors method of evaluating the
mode number, we observe significant deviations from the linear dependence of $\nu_R(M_R,\mu_R)$ on
$M_R$ when this fifth value of $M_R$ is included.
Because of this, we decided not to include the results at $M_R\approx130$ MeV.

\begin{table}[t!]
  \centering
  \begin{tabular}[]{ccc}
    Ensemble & $a\mu$ & $r_0\Sigma^{1/3}$\\
\hline
  A$30.32 $ & 0.0030 & 0.7914(53)(45)(57)\\
  A$40.32$ & 0.0040 & 0.7994(33)(45)(58)\\
 A$50.32$ & 0.0050 & 0.8120(25)(46)(59)\\
 A$60.24$ & 0.0060 & 0.8097(62)(46)(59)\\
  A$80.24$ & 0.0080 & 0.8229(38)(46)(60)\\
 chiral & & 0.7772(61)(44)(56)(157)\\
\hline
  B$25.32$ & 0.0025 & 0.7552(53)(25)(54)\\
  B$35.32$ & 0.0035 & 0.7559(35)(25)(54)\\
  B$55.32$ & 0.0055 & 0.7606(25)(26)(55)\\
  B$75.32$ & 0.0075 & 0.7730(26)(26)(56)\\
chiral & & 0.7408(55)(25)(53)(112)\\
\hline
  D$15.48$ & 0.0015 & 0.7294(44)(14)(56)\\
  D$20.48$ & 0.0020 & 0.7417(30)(14)(57)\\
  D$30.48$ & 0.0030 & 0.7425(24)(14)(57)\\
chiral & & 0.7262(72)(14)(56)(75)\\
  \end{tabular}
  \caption{Results for $r_0\Sigma^{1/3}$ for all considered $N_f=2+1+1$ ensembles. The given errors
are: statistical, resulting from $Z_P$, resulting from $r_0/a$, respectively. We also show results in
the chiral limit, where we also give the systematic error from the choice of the fitting interval (4th
error). See text for more details.}
  \label{tab:results_Nf211}
\end{table}

\begin{figure}[t!]
  \centering
  \includegraphics[width=0.7\textwidth,angle=270]{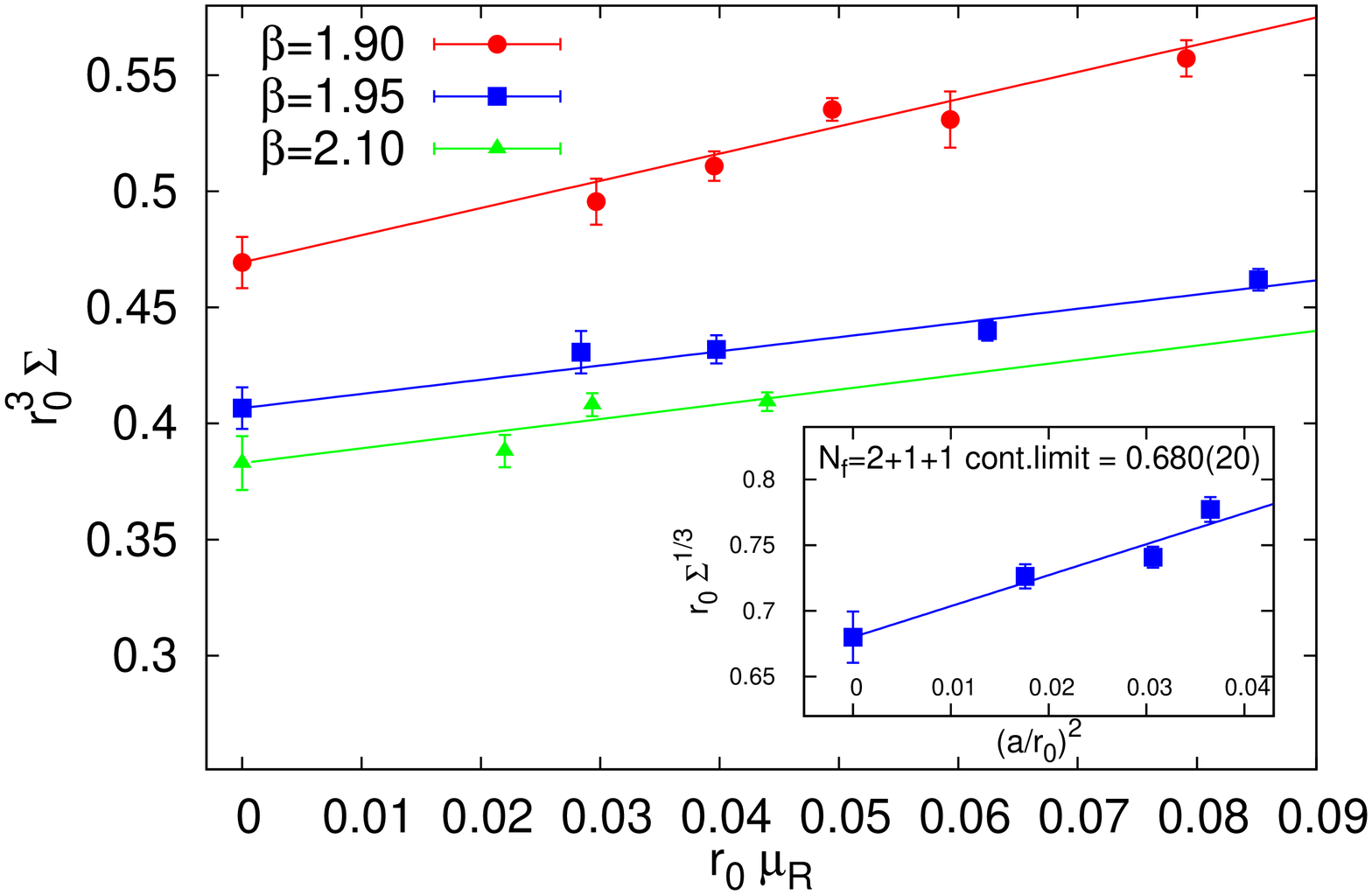}
  \caption{Main plot: chiral extrapolations of the chiral condensate $r_0^3\Sigma$ for 
    $N_f=2+1+1$ ensembles and $\beta=1.9$, 1.95, 2.1. The lines are extrapolations to the
chiral limit, linear in the quark mass. The errors are statistical only. Inset: continuum
extrapolation of the chirally extrapolated chiral condensate $r_0\Sigma^{1/3}$ vs. $(a/r_0)^2$. The
errors include: statistical errors, errors from $Z_P$ and errors from $r_0/a$.}
 \label{fig:chiral_Nf211}
\end{figure}

In Tab.~\ref{tab:results_Nf211}, we show all our results for $r_0\Sigma^{1/3}$ in the 2+1+1-flavour
case.
We also include the results of a linear extrapolation to the chiral limit for each value of $\beta$,
shown in the main plot of Fig.~\ref{fig:chiral_Nf211}.
As before, we plot only statistical errors, since all extrapolations are performed at fixed
$\beta$ and the errors from $Z_P$ and $r_0/a$ are the same for all quark masses (we use
chirally extrapolated values of $Z_P$ and $r_0/a$) -- given in Tab.~\ref{tab:results_Nf211}.
Contrary to the $N_f=2$ case, the slope of the dependence of the
mass-dependent condensate on the light quark mass slightly decreases for increasing $\beta$
(the change of slope is statistically significant when going from $\beta=1.9$ to
$\beta=2.1$).
This has the effect of lowering the systematic error related to the choice of the fitting range for
decreasing lattice spacing\footnote{Moreover, we always have at least 3 sea quark masses for each
$\beta$ in the $N_f=2+1+1$ case, compared to only 2 masses at $\beta=4.2$ ($N_f=2$)).}, which is for
all $\beta$ the dominating source of error (although for $\beta=2.1$ other errors become
comparable).

The chirally extrapolated values at three lattice spacings are then used to perform an extrapolation
to the continuum limit, which is again compatible with $\mathcal{O}(a^2)$ cut-off effects.
To estimate the fitting range uncertainty, we again perform 3 separate continuum limit extrapolations,
using different fitting ranges and different values of $M_R$, corresponding to approx. 80, 90 and 100
MeV. The values of $r_0\Sigma^{1/3}$ in the continuum limit are, respectively, 0.668(24), 0.680(20)
and 0.659(27).
As our central value we take the result from the full fitting range:
\begin{displaymath}
 r_0\Sigma^{1/3}_{N_f=2+1+1}=0.680(20)(21),
\end{displaymath}
with the larger of the differences with respect to values from tailored fitting intervals as the
fitting range systematic error.
This can be compared to the 2-flavour result which amounts to $r_0\Sigma^{1/3}_{N_f=2}=0.689(16)(29)$
and both results are compatible within errors.

\section{Conclusions}
In this paper, we presented our results on the chiral condensate in QCD with
$N_f=2$ and $N_f=2+1+1$ flavours of dynamical Wilson twisted mass quarks at maximal twist.

Our final results are:
\begin{displaymath}
 r_0\Sigma^{1/3}_{N_f=2}=0.689(16)(29),
\end{displaymath}
\begin{displaymath}
 r_0\Sigma^{1/3}_{N_f=2+1+1}=0.680(20)(21),
\end{displaymath}
which indicates that at the current level of precision, we cannot discriminate the influence of the
dynamical strange and charm quarks on the value of the light quark chiral condensate.

The main source of the error of our results is the systematic error related to the choice of the
fitting range in the dependence of the renormalized mode number on the renormalized spectral
threshold. The second most important source of the error is either the statistical error or the
error
related to the uncertainty in the values of $r_0/a$ (which are inputs of our analysis). The error
from $Z_P$ (also an input of our analysis), used to renormalize the quark masses and the spectral
threshold parameter, is usually the smallest. 
However, we want to emphasize that in all cases the fitting range error is the largest one and in
most cases it is larger by a factor of 2-4 than any other error.
The rather small statistical errors that we obtain indicate that increasing statistics would not make
our total error significantly smaller.
This implies that a way to improve the total error would be to increase the number of values of the
spectral threshold $M_R$ at which one computes the mode number.
This would allow to identify more precisely the linear region of the mode number vs. $M_R$ dependence
(see discussion in the Appendix)
-- on the one hand sufficiently far away from the renormalized quark mass, on the other hand low
enough such that there are no deviations from the linear behaviour at the upper end of the fitting
range (observed already at $M_R\approx130$ MeV in the $N_f=2+1+1$ case).

\begin{table}[t!]
  \centering
  \begin{tabular}[]{ccccc}
    Result & method & $N_f$ & fermions & $r_0\Sigma^{1/3}$\\
\hline
this work & spectral proj. & 2 & twisted mass & 0.689(16)(29)\\
this work & spectral proj. & 2+1+1 & twisted mass & 0.680(20)(21)\\
\hline
RBC-UKQCD \cite{Aoki:2010dy} & chiral fits & 2+1 & domain wall & 0.632(15)(12)\\
MILC \cite{Bazavov:2009bb} & chiral fits & 2+1 & staggered & 0.654(14)(18)\\
MILC \cite{Bazavov:2010yq} & chiral fits & 2+1 & staggered & 0.653(18)(11)\\
S. Borsanyi et al. \cite{Borsanyi:2012zv} & chiral fits & 2+1 & staggered & 0.662(5)(20)\\
ETMC \cite{Baron:2009wt} & chiral fits & 2 & twisted mass & 0.575(14)(52)\\
ETMC \cite{Burger:2012ti} & quark propagator & 2 & twisted mass & 0.676(89)(14)\\
HPQCD \cite{McNeile:2012xh} & quark propagator & 2+1+1 & staggered & 0.673(5)(11)\\
\end{tabular}
  \caption{Comparison of our results for $r_0\Sigma^{1/3}$ with large-volume continuum limit results
found in literature.
In the given references, the values of $\Sigma$ are given in MeV. To convert to
the dimensionless product $r_0\Sigma^{1/3}$, we combine them with results for $r_0$.
The first error given is always from the computation of the value in MeV (when there are several
errors given, we combine them in quadrature) and the
second one from conversion using physical value of $r_0$.
For RBC-UKQCD, $\Sigma^{1/3}=256(6)$ MeV, $r_0=0.487(9)$ fm \cite{Aoki:2010dy}.
For MILC \cite{Bazavov:2009bb}, $\Sigma^{1/3}=278(6)$ MeV, $r_1=0.318(7)$ fm \cite{Bazavov:2009bb},
$r_0/r_1$=1.46(1)(2) \cite{Bernard:2007ps}.
For MILC \cite{Bazavov:2010yq}, $\Sigma^{1/3}=281.5(7.9)$ MeV, $r_1=0.3133(23)$ fm
\cite{Bazavov:2010yq}, $r_0/r_1$=1.46(1)(2) \cite{Bernard:2007ps}.
For S. Borsanyi et al. \cite{Borsanyi:2012zv}, $\Sigma^{1/3}=272.3(1.2)(1.4)$ MeV, $r_0=0.48(1)(1)$
fm \cite{Aoki:2009sc}.
For ETMC \cite{Baron:2009wt}, $\Sigma^{1/3}=269.9(6.5)$ MeV, $r_0=0.420(14)$ fm.
However, newer analyses indicate a higher value of $r_0\approx0.45$ fm
\cite{Blossier:2010cr}. To take this into account, we added the spread of the new and old value as a
systematic error and used $r_0=0.420(38)$ fm to calculate $r_0\Sigma^{1/3}$ from $\Sigma^{1/3}$ in
MeV.
For ETMC \cite{Burger:2012ti}, $\Sigma^{1/3}=299(26)(29)$ MeV, $r_0=0.446(9)$ fm
\cite{Baron:2009wt}.
For HPQCD \cite{McNeile:2012xh}, $\Sigma^{1/3}=283(2)$ MeV, $r_1=0.3209(26)$ fm
\cite{Dowdall:2011wh}, $r_0/r_1$=1.46(1)(2) \cite{Bernard:2007ps}.
}
  \label{tab:comparison}
\end{table}

In order to place our values of the chiral condensate in context of 
results of other collaborations, we attempt 
in Tab.~\ref{tab:comparison} a 
comparison. Given the large amount of approaches to compute
the chiral condensate, as mentioned in the introduction, we make a selection 
by only considering results that are given in the literature as
continuum limit values from large volume simulations. 
Note that the other available results that are listed 
in Tab.~\ref{tab:comparison} are obtained in a different way than the spectral 
projector method. They  
mostly use chiral
perturbation theory fits
to the quark mass dependence of light pseudoscalar meson observables or
determine the chiral condensate from the quark propagator.
Not discussing the advantages or disadvantages of 
the various fermion discretizations 
used, we see in Tab.~\ref{tab:comparison} an overall agreement 
for the dimensionless quantity $r_0\Sigma^{1/3}$, which is reassuring and 
establishes, in our opinion, the spectral projector method as a valuable 
alternative to determine the chiral condensate.
On the other hand, when looking at the chiral condensate in physical units, 
see the caption of Tab.~\ref{tab:comparison}, a spread of results 
is obtained. Thus, it seems that the scale setting from the different 
lattice calculations introduces a systematic effect and it would be desirable 
to clarify this uncertainty in future more precise calculations.

\vspace{0.3cm}
\noindent {\bf Acknowledgments} We thank the European
Twisted Mass Collaboration for generating gauge field configurations ensembles used for this work.
We are grateful to A. Shindler for discussions and in particular suggestions concerning
$\mathcal{O}(a)$ improvement of the mode number.
We thank G. Herdoiza for discussions at various stages of this work and for his careful reading of the
manuscript.
We acknowledge useful discussions with P. Damgaard, V. Drach, M. L\"uscher,
K. Ottnad, G.C. Rossi, S. Sharpe, C. Urbach, F. Zimmermann.
K.C. was supported by Foundation for Polish Science fellowship ``Kolumb''.
This work was supported in part by the DFG Sonderforschungsbereich/Transregio SFB/TR9. 
K.J. was supported in part by the Cyprus Research Promotion
Foundation under contract $\Pi$PO$\Sigma$E$\Lambda$KY$\Sigma$H/EM$\Pi$EIPO$\Sigma$/0311/16.
The computer time for this project was made available to us by the J\"ulich
Supercomputing Center, LRZ in
Munich, the PC cluster in Zeuthen, Poznan Supercomputing and Networking Center (PCSS). We thank these
computer centers and their staff for all technical advice and help.

\appendix

\section{Testing the method}
\label{app:tests}

\begin{figure}[t!]
\begin{minipage}[b]{0.5\linewidth}
\includegraphics[width=1\textwidth]{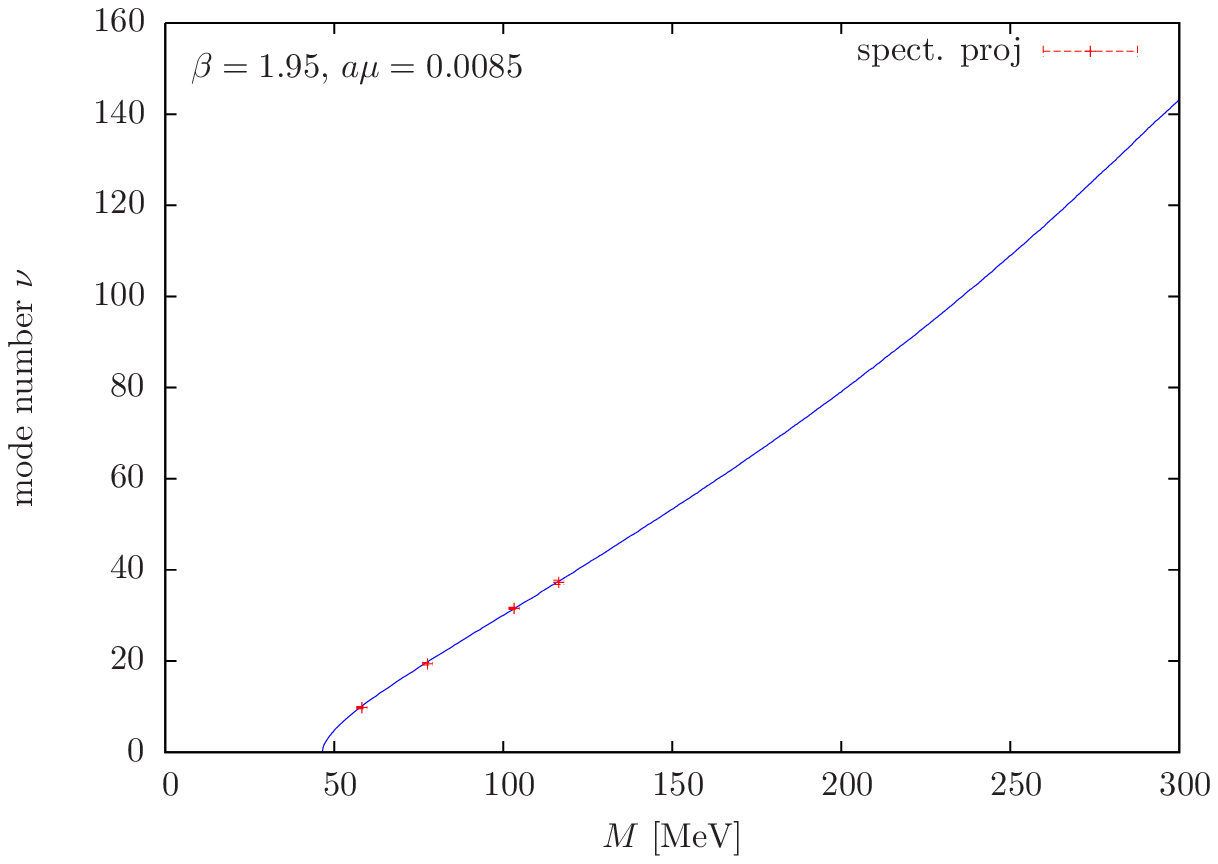}    
\label{evlstest_all}
\end{minipage}
\begin{minipage}[b]{0.5\linewidth}
\includegraphics[width=1\textwidth]{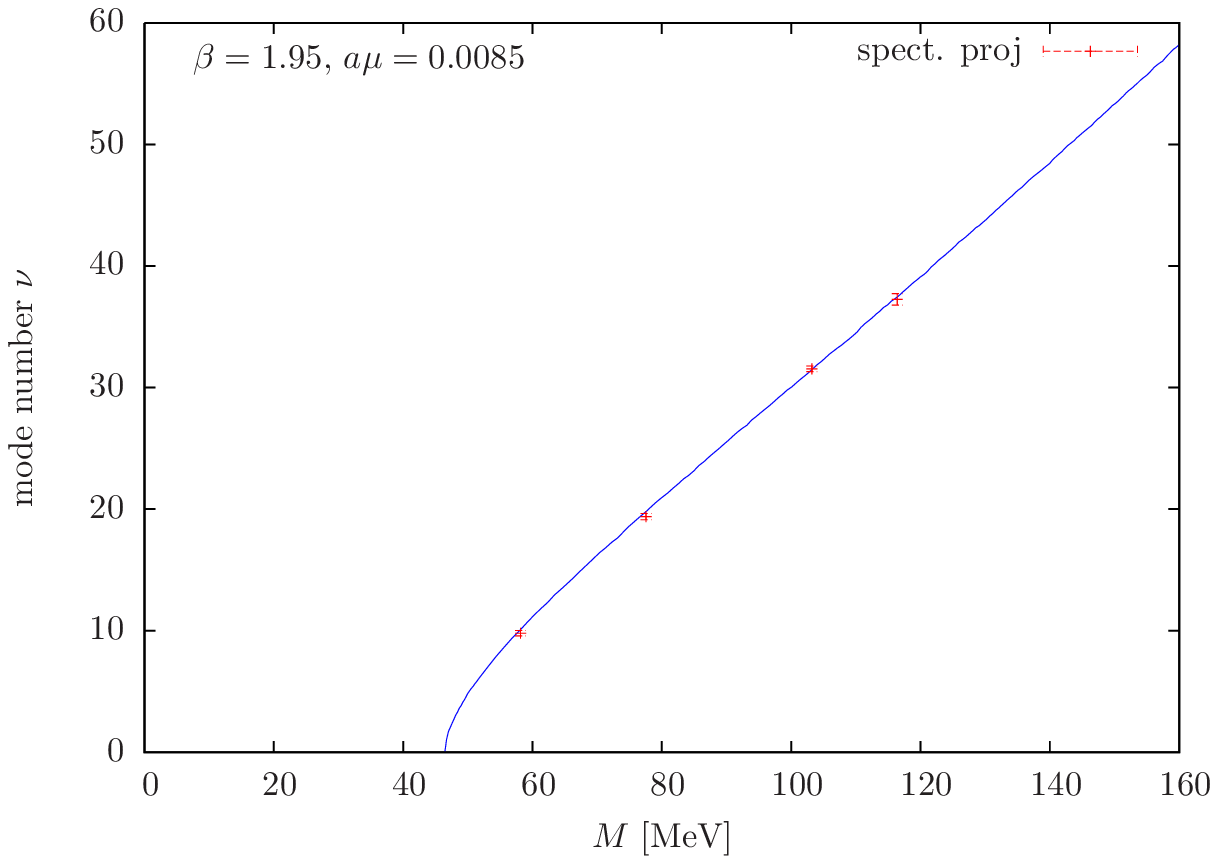}
\label{evltest_zoom}
\end{minipage}
\caption{The mode number as a function of $M_R$ for the ensemble B85.24. The solid line in both
plots is the result of explicit computation of 150 eigenvalues for each gauge field configuration,
while data points represent results of spectral projectors calculation of
the mode number for 4 values of $M_R$. The right plot is the zoom of the left part.}
\label{evltest}
\end{figure}

To test our implementation of the method in the tmLQCD package \cite{Jansen:2009xp}, we compared the
spectral projectors results for the mode number with ones from explicit computation of 150 lowest
eigenvalues of $D^\dagger D$ for each gauge field configuration, using ensemble B85.24. The results of
this comparison are shown in the right plot of Fig.~\ref{evltest}, 
where the 4 points correspond to
the stochastically evaluated mode number using spectral projectors, while the continuous line (which
has an error roughly of the order of the width of the line) is the result from explicitly computing
the eigenvalues. We observe very good agreement between the two methods.

Moreover, we used the results of explicit computation of eigenvalues to estimate the region of
renormalized spectral threshold of $M_R$ where we observe linear dependence between the renormalized
mode number and the threshold value (left plot of Fig.~\ref{evltest}). The onset of non-linear
behaviour corresponds to approx. 130-150 MeV. On the other end of the spectrum, one clearly observes
effects of $M_R$ close to the renormalized quark mass up to around 10-20 MeV above the
latter\footnote{One expects
that the behaviour close to the renormalized quark mass is modified in an important way by lattice
artefacts \cite{Necco:2011vx}.}. This
allows us to identify the linear region to extend between around 60 and 120 MeV, to allow for some
safety margin. Therefore, we decided to choose our values of $M_R$ for the computation of the chiral
condensate roughly in this interval. We remark that values in this range were used in
Ref.~\cite{Giusti:2008vb}.

Some parameters employed in the method of spectral projectors need to be tuned to obtain a compromise
between the accuracy of results and computational cost.

\begin{figure}[t!]
  \centering
\includegraphics[width=0.65\textwidth]{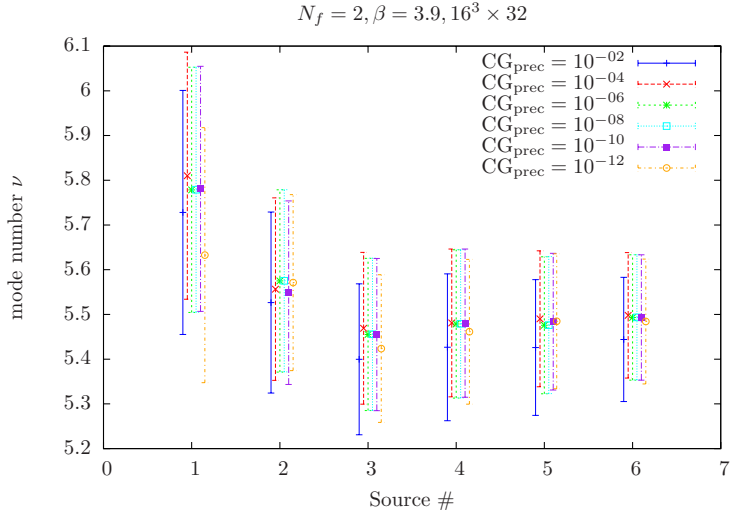}    
\caption{The mode number vs. the number of averaged stochastic sources using
  different values of the inverter relative precision for a
  $16^3\times32$ lattice at $\beta=3.9$, $a\mu=0.004$ (ensemble b40.16).}
  \label{testprec}
\end{figure}

First of all, as mentioned in Ref.~\cite{Giusti:2008vb}, the precision of the inverter can
be chosen to be relatively sloppy without reducing the accuracy.
In order to identify the optimal precision, which does not affect the
correctness of the result, but still decreases the computational time, we
computed the mode number for several values of the relative precision of the inverter.
The results (for ensemble b40.16) are shown in Fig.~\ref{testprec}, which shows that even precision of
$10^{-2}$ gives 
 reasonable result. However, we decided to be conservative and
 we chose a value of $10^{-6}$ for the relative precision of the inverter.  

We also checked the dependence of the mode number on the number of stochastic sources used for
each gauge field configuration, shown again in Fig.~\ref{testprec}.
We observe that all results are compatible within error, which matches the suggestion of
Ref.~\cite{Giusti:2008vb} that one stochastic source should be enough.
However, we observe that adding a second source might help to considerably reduce the statistical
error, which may be important for shorter Monte Carlo runs, when the available number of independent
gauge field configurations is rather small.
Adding further sources does not change the error considerably, because of correlations between
results obtained from the same gauge field configuration.

\section{Autocorrelations}
\label{sec:tauint}

\begin{table}[t!]
  \centering
  \begin{tabular}[]{ccccccc}
    \multirow{2}{*}{Ensemble} & number & step & $\tau_{\rm int}$ (boot) & $\tau_{\rm int}$ (UW)&
$\tau_{\rm int}$ (boot) & $\tau_{\rm int}$ (UW)\\
& of confs & HMC traj.& \multicolumn{2}{c}{smallest $M$} & \multicolumn{2}{c}{largest $M$}\\
\hline
  A30.32  & 116 & 20 & 0.7 & 0.6(2) & 1.7 & 1.9(8)\\
  A40.20 & 197 & 16 &0.8 & 0.9(3) & 0.8 & 0.9(3) \\
  A40.24 & 185 & 20 & 0.6 & 0.7(2) & 1.5 & 2.2(8) \\
  A40.32 & 201 & 8 & 0.8 & 0.7(2) & 0.8 & 0.7(2)\\
  A50.32 & 201 & 20 & 0.5 & 0.7(2) & 0.5 & 0.8(2)\\
  A60.24 & 161 & 8 & 0.7 & 0.5(1) & 1.3 & 1.6(6)\\
  A80.24 & 200 & 8 & 0.6 & 0.7(2) & 0.7 & 0.9(2)\\
  B25.32 & 200 & 20 & 1.0 & 0.7(2) & 1.0 & 1.2(4)\\
  B35.32 & 199 & 20 & 0.4 & 0.5(1) & 1.0 & 1.7(6)\\
  B55.32 & 201 & 20 & 0.4 & 0.4(1) & 0.6 & 0.7(2)\\
  B75.32 & 199 & 8 & 0.7 & 0.6(1) & 0.5 & 0.5(1)\\
  B85.24 & 196 & 20 & 0.4 & 0.4(1) & 0.5 & 0.5(1)\\
  D15.48 & 119 & 20 & 0.5 & 0.5(1) & 0.6 & 0.8(2)\\
  D20.48 & 193 & 20 & 0.6 & 0.5(1) & 0.7 & 0.7(2)\\
  D30.48 & 203 & 20 & 0.4 & 0.4(1) & 0.5 & 0.4(1)\\
\hline
\end{tabular}
  \caption{Autocorrelations in the mode number, $N_f=2+1+1$ ensembles. We give the number of gauge
field configurations used for each ensemble, the step in units of HMC trajectories and the calculated
values of $\tau_{\rm int}$ using two methods: bootstrap with blocking (boot) and the method proposed
by U. Wolff \cite{Wolff:2003sm} (UW).}
  \label{tab:tauint_Nf211}
\end{table}

\begin{table}[t!]
  \centering
  \begin{tabular}[]{ccccccc}
    \multirow{2}{*}{Ensemble} & number & step & $\tau_{\rm int}$ (boot) & $\tau_{\rm int}$ (UW)&
$\tau_{\rm int}$ (boot) & $\tau_{\rm int}$ (UW)\\
& of confs & HMC traj.& \multicolumn{2}{c}{smallest $M$} & \multicolumn{2}{c}{largest $M$}\\
\hline
  b$30.32 $ & 134 & 8 & 0.3 & 0.4(1) & 0.9 & 0.9(3)\\
   b$40.16 $  & 544 & 10 & 0.5 & 0.5(1) & 1.6 & 1.8(4)\\
  b$40.20 $  & 265 & 20 & 0.6 & 0.5(1) & 1.2 & 1.3(3)\\
  b$40.24 $  & 465 & 20 & 0.5 & 0.7(2) & 0.8 & 1.3(3)\\
  b$40.32$ &  232 & 16 & 0.4 & 0.4(1) & 0.5 & 0.5(1)\\
 b$64.24$ & 272 & 20 & 0.4 & 0.4(1) & 0.6 & 0.8(2)\\
 b$85.24$ & 187 & 20 & 0.3 & 0.4(1) & 0.5 & 0.5(1)\\
  c$30.32$ & 183 & 8 & 0.7 & 0.5(1) & 1.2 & 1.6(5)\\
  c$60.32$ & 123 & 20 & 0.5 & 0.5(1) & 0.4 & 0.5(1)\\
  c$80.32$ & 201 & 10 & 0.4 & 0.5(1) & 1.1 & 0.8(2)\\
  d$20.48$ & 77 & 10 & 0.7 & 0.5(1) & 0.5 & 0.6(2)\\
  d$65.32$ & 199 & 20 & 0.6 & 0.5(1) & 0.7 & 1.0(3)\\
\hline
\end{tabular}
  \caption{Autocorrelations in the mode number, $N_f=2$ ensembles. We give the number of gauge
field configurations used for each ensemble, the step in units of HMC trajectories and the calculated
values of $\tau_{\rm int}$ using two methods: bootstrap with blocking (boot) and the method proposed
by U. Wolff \cite{Wolff:2003sm} (UW).}
  \label{tab:tauint_Nf2}
\end{table}

In this appendix, we show the results of our autocorrelation analysis of the mode number (for the
smallest and largest value of $M$ that we used for chiral condensate extraction).
We applied two methods: bootstrap with blocking (with block size of 10 measurements) and the method
proposed by U.~Wolff in Ref.~\cite{Wolff:2003sm}.
Our results are shown in Tabs.~\ref{tab:tauint_Nf211} ($N_f=2+1+1$) and \ref{tab:tauint_Nf2}
($N_f=2$).
In general, both methods yield compatible results for the integrated autocorrelation time $\tau_{\rm
int}$ (in the case of the method of Ref.~\cite{Wolff:2003sm}, we quote also the error of $\tau_{\rm
int}$ and the two values of $\tau_{\rm int}$ that we obtain agree within this error).

Our conclusions about the dependence of $\tau_{\rm int}$ on simulation parameters are the following:
\begin{itemize}
 \item at the smallest value of $M$, no autocorrelations are observed ($\tau_{\rm int}$ compatible
with 0.5),
 \item at the largest value of $M$, in some cases the autocorrelations become visible, with $\tau_{\rm
int}$ between 1 and 2,
 \item there is a tendency towards larger autocorrelations for smaller quark masses (e.g. B25
compared to B85) and for smaller volumes (e.g. b40 ensembles at 4 volumes),
 \item we don't observe a tendency towards increased $\tau_{\rm int}$ for decreasing lattice spacing.
\end{itemize}

\bibliographystyle{jhep}
\bibliography{ch_cond}	

\end{document}